\documentclass[runningheads]{llncs}
\usepackage[T1]{fontenc}
\usepackage{graphicx}
\usepackage{xspace}
\usepackage{makecell}
\usepackage[hidelinks]{hyperref}

\hypersetup{
    colorlinks=false,
    linkcolor=blue,
    filecolor=magenta,      
    urlcolor=cyan,
    pdftitle={Overleaf Example},
    pdfpagemode=FullScreen,
}

\usepackage{soul}
\usepackage{xcolor}

\newcommand{\msmarco}{\textsc{MsMarco}\xspace}
\newcommand{\msmarcodos}{\textsc{MsMarco v2}\xspace}

\newcommand{\seismic}{\textsc{Seismic}\xspace}

\newcommand{\hnsw}{\textsc{Hnsw}\xspace}

\newcommand{\kannolo}{\textsc{kANNolo}\xspace}

\newcommand{\grassrma}{\textsc{GrassRMA}\xspace}
\newcommand{\pyann}{\textsc{PyAnn}\xspace}

\newcommand{\splade}{\textsc{Splade}\xspace}

\newcommand{\cut}{\textsf{cut}}
\newcommand{\heapfactor}{\textsf{heap\_factor}\xspace}

\newcommand{\etal}{\emph{et al.}\xspace}

\newcommand{\knn}{$\kappa$-NN\xspace}

\newcommand{\bigann}{\textsc{BigAnn}\xspace}

\usepackage[show]{chato-notes}
\usepackage{adjustbox}
\usepackage{booktabs}
\usepackage{enumitem}
\usepackage{fontawesome5}

\begin{document}
\title{Investigating the Scalability of Approximate Sparse Retrieval Algorithms to Massive Datasets}
\titlerunning{Scalability of Approximate Sparse Retrieval Algorithms to Massive Datasets}
%
\author{Sebastian Bruch\inst{1}$^\star$\orcidID{0000-0002-2469-8242
}\and \\ 
Franco Maria Nardini\inst{2}$^\star$\orcidID{0000-0003-3183-334X}\and \\  
Cosimo Rulli\inst{2}$^\star$\faIcon{envelope}\orcidID{0000-0003-0194-361X} \and \\
Rossano Venturini\inst{3}$^\star$\orcidID{0000-0002-9830-3936} \and \\ 
Leonardo Venuta\inst{4}\thanks{All authors contributed equally to this work.}
}

\authorrunning{Bruch \emph{et al.}}
%
\institute{Northeastern University, Boston, USA,
\email{s.bruch@northeastern.edu}
\and
ISTI-CNR, Pisa, Italy,
\email{\{name.surname\}@isti.cnr.it}\\
\and
University of Pisa, Italy,
\email{rossano.venturini@unipi.it}
\and
University of Pisa, Italy,
\email{l.venuta@studenti.unipi.it}
}
\maketitle
%
\begin{abstract}
Learned sparse text embeddings have gained popularity due to their effectiveness in top-$k$ retrieval and inherent interpretability. Their distributional idiosyncrasies, however, have long hindered their use in real-world retrieval systems. That changed with the recent development of approximate algorithms that leverage the distributional properties of sparse embeddings to speed up retrieval. Nonetheless, in much of the existing literature, evaluation has been limited to datasets with only a few million documents such as \textsc{MsMarco}. It remains unclear how these systems behave on much larger datasets and what challenges lurk in larger scales. To bridge that gap, we investigate the behavior of state-of-the-art retrieval algorithms on massive datasets. We compare and contrast the recently-proposed \textsc{Seismic} and graph-based solutions adapted from dense retrieval.
We extensively evaluate \textsc{Splade} embeddings of $138$M passages from \textsc{MsMarco-v2} and report indexing time and other efficiency and effectiveness metrics.

\keywords{Sparse Embeddings \and Approximate Large-scale Retrieval.}
\end{abstract}

\section{Introduction}
\label{sec:intro}

Sparse embeddings~\cite{epic,splade-sigir2021,formal2021splade,formal2022splade,lassance2022efficient-splade} have proven effective in capturing the contextual semantics of text. Such embeddings have matured over the past few years and found their way into first-stage retrieval~\cite{INR-071}.

Neural models that are trained to produce sparse embeddings rewire a Large Language Model (LLM) to generate a high-dimensional vector for an input text. Each coordinate of the output represents a term from the model's vocabulary, and if it is nonzero, its corresponding term is taken to be  \emph{semantically} relevant to the input. Similarity between embeddings is by inner product, making top-$k$ retrieval an instance of Maximum Inner Product Search (MIPS).

Sparse embeddings have three useful properties. First, repeated experiments show that they are as effective as \emph{dense embeddings}~\cite{DBLP:series/synthesis/2021LinNY,karpukhin-etal-2020-dense,xiong2021approximate,reimers-2019-sentence-bert,santhanam-etal-2022-colbertv2,colbert2020khattab,10.1007/978-3-031-56060-6_1} and generalize better to out-of-domain datasets~\cite{bruch2023fusion,lassance2022efficient-splade,thakur2023sprint}. Second, the one-to-one mapping between dimensions and terms makes them inherently \emph{interpretable}. Third, they fit naturally in the inverted index-centric paradigm~\cite{tonellotto2018survey}, while remedying the \emph{vocabulary mismatch} problem at the same time.

Unfortunately, a na\"ive application of inverted index-based retrieval to sparse embeddings does not satisfy tight latency constrains of real-world systems. That stems from distributional differences between embeddings and bag-of-words representations of text~\cite{bruch2023sinnamon,mackenzie2021wacky}. That challenge gave rise to works that attempt to speed up (safe or unsafe) top-$k$ retrieval for sparse embeddings~\cite{bruch2023sinnamon,bruch2023bridging,lassance2023static-pruning,mallia2022guided-traversal,10.1145/3576922}.

Among these methods, the approximate (unsafe) retrieval method by Bruch \etal~\cite{bruch2024seismic} stands out. The algorithm they call \seismic uses an inverted index as usual, but organizes its inverted lists into geometric blocks---similar in spirit to~\cite{Cinar2023ClusterSkipping,Altingovde2008ClusterSkipping}. Each of these blocks is then represented with a \emph{summary} vector. During query processing, \seismic judges if a block must be evaluated based on whether or not its summary has a ``high-enough'' inner product with the query.

A comprehensive empirical evaluation of \seismic on sparse embeddings of \msmarco~\cite{nguyen2016msmarco} showed that it reaches sub-millisecond per-query latency with high recall. That is one to two orders of magnitude faster than inverted index-based solutions and more efficient than graph-based methods from the $2023$ \bigann Challenge~\cite{simhadri2024resultsbigannneurips23}. Later, the same authors improved \seismic further, making it almost exact but up to $2.2\times$ faster than the original \seismic~\cite{bruch2024seismicWave}.

In all the works discussed above, the retrieval algorithms have been evaluated only on datasets no larger than \msmarco, which is made up of roughly $8.8$ million embedding vectors. What happens if we applied these methods to a much larger dataset, such as \msmarcodos, with about $138$ million embedding vectors? What surprising challenges lurk in such scales that do not ordinarily surface? How long will indexing take and how much memory will it require? Does retrieval accuracy deteriorate as the collection grows in size?

We investigate these questions empirically by subjecting approximate retrieval algorithms designed for learned sparse embeddings to this massive dataset. To the best of our knowledge, no work has yet studied the efficiency and effectiveness of such methods at the extreme scale we consider in this work.
\section{Experimental Setup}
\label{sec:setup}
In this study, we compare two main families of retrieval algorithms over sparse embeddings of massive datasets: graph-based and inverted index-based.

\vspace{1mm}
\noindent \textbf{Graph-based algorithms}. The 2023 \bigann Challenge~\cite{simhadri2024resultsbigannneurips23} organized a track for sparse vector retrieval over \splade~\cite{formal2023tois-splade} embeddings of \msmarco~\cite{nguyen2016msmarco}. Two graph methods emerged as clear winners: \pyann and \grassrma.

\pyann builds a graph index with \hnsw~\cite{hnsw2020}, but uses a modified search algorithm during query processing. In particular, it quantizes vectors so that coordinates are represented as $16$-bit integers, and values as $16$-bit half-precision floats. Furthermore, it discards smaller values of query vectors. The retrieved set is reranked in a refinement step that uses full vectors to compute inner products.

\grassrma also uses \hnsw for graph index construction. Similar to \pyann, it introduces a few minor optimizations: it co-locates coordinates and values to improve memory access, and employs an upper- and lower-bound of the values per vector to terminate inner product computation early.

Recently, a new implementation of \hnsw has been made available in the \kannolo~\cite{delfino2025kannolo} library,\footnote{\url{https://github.com/TusKANNy/kannolo}} a framework in Rust for approximate nearest neighbors search targeting both dense and sparse domains. We choose this solution because, as Delfino \emph{et al.} show, it outperforms both \grassrma and \pyann in terms of retrieval time~\cite{delfino2025kannolo}.



\vspace{1mm}
\noindent \textbf{\seismic}.
In contrast to the methods above, \seismic operates on the inverted and the forward index. The crucial innovation is that, each inverted list is clustered into geometric blocks, each equipped with a \emph{summary} vector. An example summary would be a vector that records the maximum of each nonzero coordinate among vectors in that block. See~\cite{bruch2024seismic} for the full description.

\seismic executes a term-at-a-time strategy to produce the top-$k$ results for a query $q$. When traversing an inverted list, it first computes the inner product between $q$ and every summary in that list to compute a ``potential'' score. It then visits each block, comparing its potential with the smallest score in the top-$k$ heap. If a block's potential exceeds that threshold, \seismic uses the forward index to compute the exact inner product between $q$ and every document in that block. This allows the query processor to skip over a large number of blocks.

A follow-up work~\cite{bruch2024seismicWave} adds a \knn graph to the index: a graph where vectors are nodes, and each node is connected to its $\kappa$ nearest neighbors by inner product. The updated algorithm uses the \knn graph as follows. Once the retrieval procedure described above concludes its search, it takes the set of $k$ documents in the heap, denoted by $\mathcal{S}$. It then forms the \emph{expanded set} $\tilde{\mathcal{S}} = \bigcup_{u \in \mathcal{S}} \big( \{ u \} \cup \mathcal{N}(u) \big)$, where $\mathcal{N}(u)$ is the set of neighbors of $u$ in the \knn graph; computes scores for documents in $\tilde{\mathcal{S}}$ using the forward index; and returns the top-$k$ subset. The \seismic code is available on GitHub.\footnote{\url{https://github.com/TusKANNy/seismic}}

\vspace{1mm}
\noindent \textbf{Dataset}. We conduct our experiments on \splade~\cite{splade-sigir2021} embeddings of \msmarcodos, a collection of $138$ million passages with $3{,}903$ queries in the \texttt{dev1} set.\footnote{\texttt{cocondenser-ensembledistil} version from \url{https://github.com/naver/splade}.}
Retrieval over these embeddings yields MRR@$10$ of 10.88\%, MRR@$100$ of $11.74\%$, and Recall@$1000$ of $61.87\%$.\footnote{The embeddings of \msmarcodos used in this study are made available at \url{https://huggingface.co/collections/tuskanny}.} Passages (queries) have $127$ ($44$) nonzero entries each, which is close to the statistics of \splade on \msmarco: $119$ ($43$). 

\vspace{1mm}
\noindent \textbf{Index size budget}.
We allocate memory budgets for the index as multiples of the dataset size, which is about $66$GB with $16$-bit floats for values and $16$-bit integers for components. We only consider hyperparameters that result in indexes that are up to $1.5\times$ the dataset size in one scenario, and $2\times$ in another.

\vspace{1mm}
\noindent \textbf{Hyperparameters}. The graph index has the following hyperparameters: $M$ (number of neighbors per node), and $ef_c$ (number of nodes scanned to build a node's neighborhood). We let $M \in \{16, 32, 64 \}$ and $ef_c = 500$. 
For \seismic, we set: number of postings, $\lambda \in \{3, 4, 5, 6 , 7, 8, 9\} \times 10^4$; summary energy, $\alpha=0.4$; and number of blocks per list, $\beta = \lambda/10$. These are the best reported values on \msmarco. We let $\kappa$ to take values in $\{10, 20\}$, and
build an (approximate) \knn graph by using a \seismic index with $\lambda = 6 \times 10^4$, \cut{} set to $20$, and \heapfactor to $0.6$.

We pick the best configuration following the same protocol as~\cite{bruch2024seismic,bruch2024seismicWave,simhadri2024resultsbigannneurips23}. We reiterate that, if our budget is $1.5\times$ the dataset size, we limit our search to hyperparameters that produce an index that is smaller than that budget.

Given an index, we process queries using the following hyperparameters and evaluate each configuration separately.
For \seismic, we vary \cut \ in $\{2, .., 14\}$, with step 2, \heapfactor in $\{ 0.6, .. , 1.0\}$, and $\kappa$ in $\{10, 20\}$. For the graph index, we vary $ef_s$ in $[10, 50]$ with step $5$, $\{50, 100\}$ with step $10$, $\{100, 1500\}$ with step $100$. For both methods, we consider the shortest time it takes the algorithm to reach accuracy@10 for values in $\{90, .., 98\}$.

\vspace{1mm}
\noindent \textbf{Metrics}. We evaluate along the following axes:
\begin{itemize}[leftmargin=*]
    \item Average Latency (milliseconds, $msec$): Elapsed time to retrieve top-$k$ vectors for a query with a single thread. This does not include embedding time.
    \item Accuracy: Percentage of true nearest neighbors recalled.
    \item Index size (GB): The space the index occupies in memory.
    \item Indexing time: The time required to build the index with all available threads.
\end{itemize}

\vspace{1mm}
\noindent \textbf{Hardware Details}.
We use version $1{.}81$ of the Rust compiler and compile using the \texttt{release} option. We run experiments on a NUMA server
with $1$TiB of RAM and four Intel Xeon Gold 6252N CPUs ($2{.}30$ GHz), with a total of $192$ cores ($96$ physical and $96$ hyper-threaded). Using the \texttt{numactl}\footnote{\url{https://linux.die.net/man/8/numactl}} tool, we enforce that the execution of the retrieval algorithms is done on a single CPU and its local memory. In this way, we avoid performance degradation due to non-uniform memory accesses across different CPUs. We note that, each CPU node has enough local memory ($256$GiB) to store entire indexes. We leave the study of cases where the index does not fit in the main memory to future work.

\section{Experimental Results}
\label{sec:exp}

\begin{figure}[!t]
    \centering
    \includegraphics[width=0.85\linewidth]{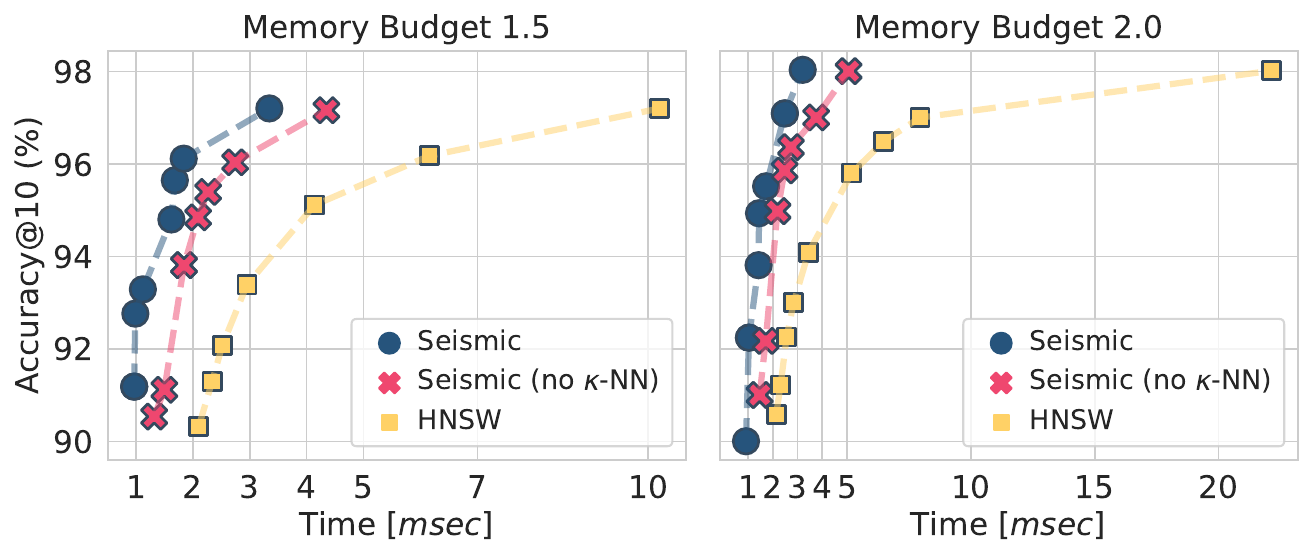}
    \caption{Comparison of \hnsw and \seismic (with and without \knn graph) by accuracy at $k=10$ as a function of query latency. We allow hyperparameters that result in an index whose size is at most $1.5\times$ (left) or $2\times$ (right) the size of the dataset. }
    \label{fig:k10}
\end{figure}

Our first experiments, visualized in Figure~\ref{fig:k10}, test \hnsw as implemented in \kannolo and \seismic by accuracy@$10$ as a function of query latency.
\seismic remains Pareto-efficient relative to \hnsw in both memory budget settings.

Interestingly, when we allow the algorithms to use more memory (i.e., $2\times$ the dataset size), the gap between \seismic and \hnsw widens as accuracy increases. That is thanks to the \knn graph, which gives an advantage in high accuracy scenarios ($>=97\%$), as validated by a comparison of \seismic with and without the \knn graph. As an example, in the $1.5\times$ scenario, \seismic with (without) \knn is $3.0\times$ ($2.3\times$) faster than \hnsw at 97\% accuracy cutoff.  The speedups increase to $6.9\times$ ($4.4\times$) in the $2\times$ scenario at 98\% accuracy cutoff.

Observe that the \knn graph has a notable impact on memory: Storing it costs $ (\lfloor \log_{2} (n-1)\rfloor  + 1)n\kappa$ bits, with $n$ denoting the size of the collection, which translates to $27$ bits per document in our setup. This analysis suggests that reducing the memory impact of the \knn graph, for example, using delta encoding to store the ids of the documents, may help \seismic become even faster at a given memory budget.

\vspace{1mm}
\noindent \textbf{Index construction time and size}.
\seismic builds the \msmarco index much more quickly than \hnsw (c.f., Table~$2$ in Bruch ~\emph{et al.}~\cite{bruch2024seismic}). This advantage holds strong on \msmarcodos, as we show in Table~\ref{table:indexing-time}: \seismic (without the \knn graph) builds its index in about $30$ minutes, while \hnsw in the sparse domain takes about $12.5$ hours. The reported configuration refers to the $1.5\times$ memory budget scenario.

Enabling the \knn graph in \seismic substantially increases the indexing time.
That is to be expected: Every document in the collection becomes a query. 
Despite the construction of the \knn graph being approximated with \seismic, this process involves the daunting task of searching through approximately 140 million queries. Moreover, as the collection size grows, not only would one have more queries to search, but there would be a larger number of documents to search over. As such, the inference benefits of the \knn graph shown in Figure~\ref{fig:k10} come at a high cost at indexing time.  




    

\begin{table}[b]
\caption{Index size and build time for winning configurations.}
 \label{table:indexing-time}
	\centering
\adjustbox{max width=\textwidth}{
	\begin{tabular}{l|@{\hspace{15pt}}l@{\hspace{15pt}}r @{\hspace{15pt}}r}
    \toprule
    Model & Configuration & \thead{Index size \\(GB)}   & \thead{Build time \\ (hours)} \\
    \midrule
    \hnsw & M=32 &104.4  &  10.1  
   \\

    \seismic & $\lambda=4e4$, $\kappa =20$& 98.4 & 16.6 \\
    
    \seismic (no $\kappa$-NN) &$\lambda=6e4$ & 98.5 & 0.5\\ 
    
     \bottomrule
	 \end{tabular}
}
\end{table}


\subsection{Scaling Laws}

\begin{figure}[!t]
    \centering
    \includegraphics[width=0.85\linewidth]{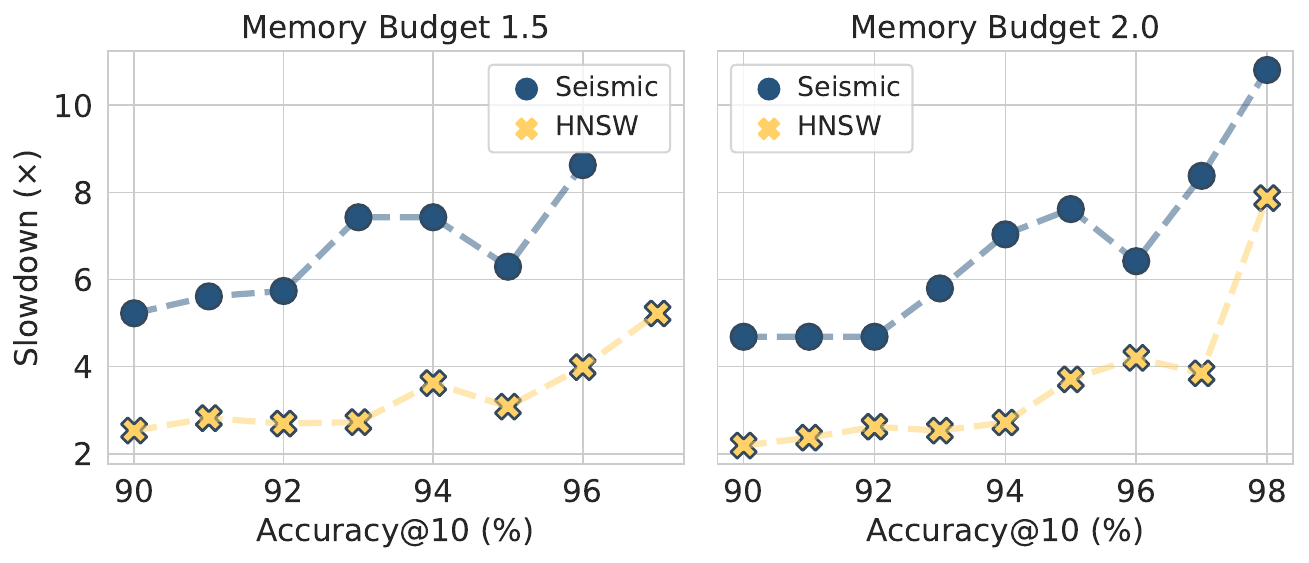}
    \caption{Scaling laws of \seismic and sparse \hnsw (as provided by the \kannolo library). For each accuracy cutoff, we measure the ratio between the latency of a method on \msmarcodos and on \msmarco.}
    \label{fig:scaling}
\end{figure}

In Figure~\ref{fig:scaling}, we present the retrieval time scaling laws for \seismic and \hnsw. For each accuracy threshold on the horizontal axis, we plot on the vertical axis the ratio of the search time on \msmarcodos to search time on \msmarco. In doing so, we report the amount of slowdown between the two datasets. In computing these ratios, we use the same hyperparameters as Bruch \emph{et al.}~\cite{bruch2024seismicWave} for \msmarco; for \msmarcodos, we use the hyperparameters noted earlier that gave the results in Figure~\ref{fig:k10}.

Considering the fact that \msmarcodos is approximately $15\times$ larger than \msmarco, from Figure~\ref{fig:scaling} we learn that both methods scale effectively with dataset size. Concerning \seismic, all configurations show a slowdown of less than $8\times$, except for accuracy thresholds of $97$ and $98$. Regarding \hnsw, in the $1.5\times$ memory budget scenario, slowdowns are reduced, ranging from $2.5\times$ to $5.2\times$. In the $2\times$ memory budget, the slowdown is small for mid accuracy cuts ($90$-$94$) and then , as for \seismic, it increases reaching a peak of $7.9\times$ at $98$.

\section{Conclusions and Future Work}
\label{sec:conclusion}
We investigated the performance of state-of-the-art approximate sparse retrieval algorithms on \splade embeddings of \msmarcodos. We focused our analysis to the recently-proposed \seismic and graph-based algorithms.

Our analysis confirms what has been known anecdotally: Building a graph index for approximate nearest neighbor search is a resource-intensive and time-consuming effort. Building an \hnsw index (using an efficient implementation from \kannolo) takes $25\times$ more time than \seismic.

Our experiments also reveal that \seismic processes queries with a much lower latency than a graph method, and that the gap between the two methods widens as we demand a higher retrieval accuracy. For example, at $98\%$ accuracy@$10$ in the $2\times$ budget setting, \seismic returns results in $3$ milliseconds per query, whereas sparse \hnsw is $6.9\times$ slower. It is, however, worth noting that \seismic's query latency scales less favorably than \hnsw.

As future work, we intend to investigate efficient retrieval over massive datasets addressing two specific scenarios: 1) efficient retrieval where indexes cannot fit completely in main memory and 2) efficient retrieval in low-resource environments (i.e., on devices with low computational resources, memory, or disk).

\section*{Acknowledgments}
This work was partially supported by the Horizon Europe RIA ``Extreme Food Risk Analytics'' (EFRA), grant agreement n. 101093026, by the PNRR - M4C2 - Investimento 1.3, Partenariato Esteso PE00000013 - ``FAIR - Future Artificial Intelligence Research'' - Spoke 1 ``Human-centered AI'' funded by the European Commission under the NextGeneration EU program, by the PNRR ECS00000017 Tuscany Health Ecosystem Spoke 6 ``Precision medicine \& personalized healthcare'' funded by the European Commission under the NextGeneration EU programme, by the MUR-PRIN 2017 ``Algorithms, Data Structures and Combinatorics for Machine Learning'', grant agreement n. 2017K7XPAN\_003, and by the MUR-PRIN 2022 ``Algorithmic Problems and Machine Learning'', grant agreement n. 20229BCXNW.


\bibliographystyle{splncs04}
\bibliography{biblio}
\end{document}